\documentclass[twocolumn,aps,prl,draft,showpacs,floatfix]{revtex4}
\usepackage{graphicx}
\usepackage{bm}
\begin{document}
\title{Scaling and universality in turbulent convection}
\author{Antonio Celani$^1$, Takeshi Matsumoto$^{2,3}$, Andrea Mazzino$^{4,5}$,
and Massimo Vergassola$^2$}
\affiliation{$^1$ CNRS, INLN, 1361 Route des Lucioles, 06560 Valbonne, 
France.\\
$^2$CNRS, Observatoire de la C\^ote d'Azur, B.P. 4229,
06304 Nice Cedex 4, France.\\
$^3$Dept. of Physics, Kyoto University, Kyoto 606-8502, Japan.\\
$^4$CNR--ISIAtA, Polo Scientifico dell Universit\`a, 
I--73100, Lecce, Italy.\\
$^5$INFM--Dipartimento di Fisica, Universit\`a di Genova,
Via Dodecaneso, 33, I-16146 Genova, Italy.}
\date{\today}
\begin{abstract}
Anomalous correlation functions of the temperature field 
in two-dimensional turbulent convection 
are shown to be universal with respect to the choice of external sources.
Moreover, they are equal to the anomalous correlations 
of the concentration field of a passive tracer
advected by the convective flow itself. The statistics of velocity differences
is found to be universal, self-similar and close to Gaussian. 
These results point to the conclusion that 
temperature intermittency in two-dimensional turbulent convection 
may be traced back to the existence of statistically preserved structures,
as it is in passive scalar turbulence.
\end{abstract}
\maketitle
\begin{figure}[hb]
\includegraphics[draft=false, scale=0.6]{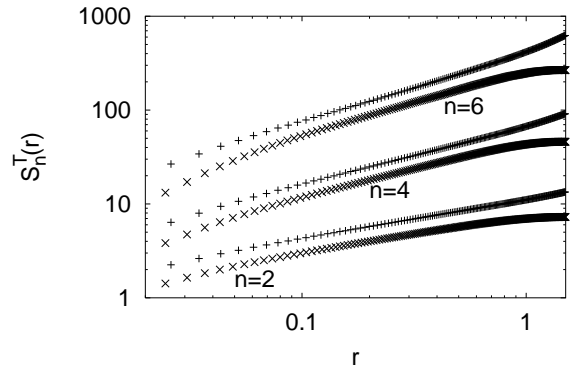}
\caption{The moments of 
temperature differences, $S^T_n(r)$, for $n=2,4,6$,
as a function of the separation $r$.  Note the parallelism
between curves of the same order $n$, within the scaling range. 
The two sets of curves are generated by two different kinds of injection 
mechanisms. In the first case ($\times$), $f_T$ is a random Gaussian forcing,
with correlation $\langle f_T({\bm r},t)f_T({\bm r}',t')\rangle=
F_T(|{\bm r}-{\bm r}'|)\delta(t-t')$, 
where $F_T$ decays with the characteristic scale
$L$ (approximately one-fourth of the box-size); in the second case ($+$),
the system is driven by the term $f_T=\gamma{\bm g}\cdot{\bm u}$, that
mimics the effect of a mean temperature gradient on the transport
of temperature fluctuations. We consider only the isotropic contribution
to the statistics, by averaging over all directions of
the separation ${\bm r}$.
For orders equal or larger than $n=8$ all
exponents collapse -- within errorbars --  on the saturation 
value $\zeta^T_{\infty} \simeq 0.8$. 
The curves have been multiplied 
by appropriate numerical factors for viewing purposes.
The equality of the scaling exponents $\zeta^T_n$ for the two types of forcing
has been checked by computing the logarithmic slope $d \ln S^T_n(r) / \ln r$
(not shown).}
\label{fig:3} 
\end{figure}
\begin{figure}[hb]
\includegraphics[draft=false, scale=0.6]{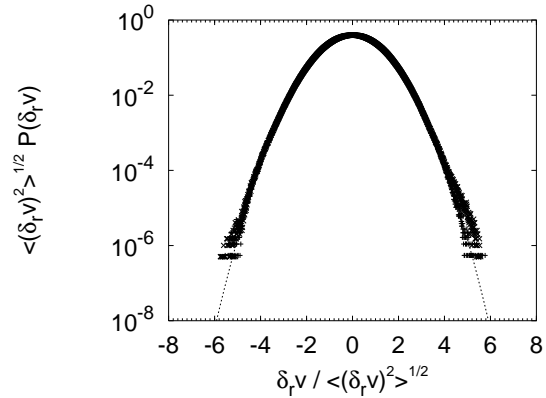}
\caption{Probability density function of longitudinal velocity increments
$\delta_r v = ({\bm v}({\bm r},t)-{\bm v}({\bm r},0))\cdot \hat{\bm r}$,
rescaled to their standard deviation $\langle (\delta_r v)^2\rangle^{1/2}$.
We show two sets of data obtained by
driving the system with random Gaussian forcing ($\times$), and by  
 $f_T=\gamma{\bm g}\cdot{\bm u}$ ($+$). Here $r=0.2$, inside the scaling range.
At different $r$ the rescaled pdf's collapse onto each other, 
as expected for a self-similar statistics. The Gaussian density function is 
 shown as a dotted line, for comparison.}
\label{fig:5} 
\end{figure}
\begin{figure}[ht]
\includegraphics[draft=false, scale=0.6]{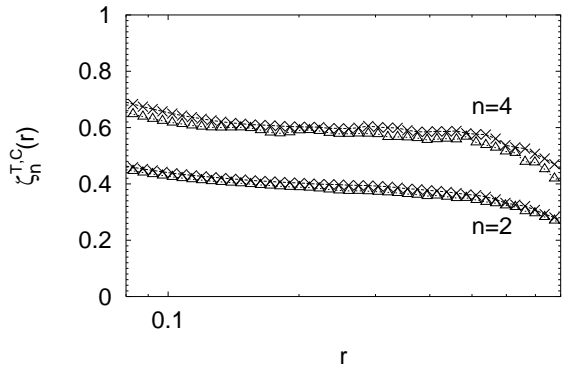}
\caption{Local scaling exponents of temperature ($\times$) 
and concentration ($\bigtriangleup$) fluctuations,
$\zeta^{T,C}_n(r) = d \ln S^{T,C}_n(r) / d \ln r$. 
Temperature and concentration are driven 
by independent Gaussian random forcings.}
\label{fig:6} 
\end{figure}

Heat and momentum transport in slightly heated flows 
are governed by the Boussinesq equations \cite{MY71} 
\begin{equation}
\begin{array}{cc} 
\partial_t T+{\bm v}\cdot {\bm \nabla} T =\kappa\Delta T + f_T \\
\partial_t {\bm v} +{\bm v}\cdot {\bm \nabla} {\bm v} = -{\bm \nabla} p 
- \beta  T  {\bm g} + \nu\Delta{\bm v} 
 \;, \end{array} 
\label{eq:1} 
\end{equation} 
where $T$ is the field of the
temperature fluctuations, 
${\bm v}$ is the velocity field,
${\bm g}$ is the gravitational acceleration, 
$\beta$ is the thermal expansion coefficient and $\kappa$, 
$\nu$ are the molecular diffusivity and viscosity. 
The system is kept in a statistically stationary
state by the external source of fluctuations $f_T$.
We will focus on the statistical
properties of temperature excursions at scales 
larger than the Bolgiano scale $l_B$ -- where buoyancy forces 
balance the inertial ones in the velocity dynamics --
yet smaller than the forcing correlation length, $L$.
In that range, temperature fluctuations cascade toward the small scales
where they are eventually dissipated by thermal diffusivity.
\begin{figure}[hb]
\centerline{\hspace{2cm}
\includegraphics[draft=false, scale=0.5, clip=true]{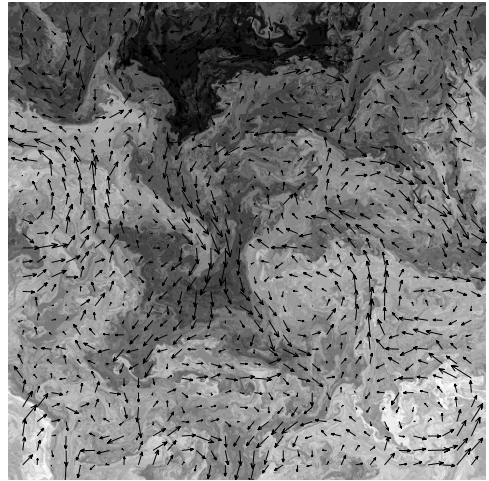} 
\vspace{0.3cm}}
\caption{Snapshot of the temperature and velocity fields. Dark areas
identify cold regions. The Boussinesq
equations~(\protect{\ref{eq:1}}) are solved in a two-dimensional 
doubly periodic domain, with $1024^2$ collocation points.
The Bolgiano scale $l_B$ is comparable to the smallest resolved length-scale. 
Since in two dimensions there is a net energy flux toward the large scales,
a statistically steady state requires 
the momentum equation in~(\protect{\ref{eq:1}}) be supplemented by a 
friction term $-\alpha {\bm v}$ that drags energy from the gravest modes. 
As customary, diffusive and viscous terms are replaced by 
hyperdiffusive ($-\kappa \Delta^2$) and hyperviscous ($-\nu \Delta^4$) ones,
to confine dissipative effects to the smallest scales.}
\label{fig:1} 
\end{figure}
\begin{figure}[hb]
\includegraphics[draft=false, scale=0.6]{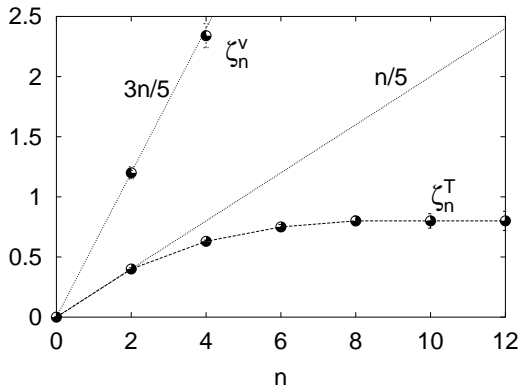}
\caption{Scaling exponents of temperature, $\zeta^T_n$,  
and velocity, $\zeta^v_n$. The straight lines are the dimensional predictions,
$n/5$ for temperature, $3n/5$ for velocity. Notice that at orders larger
than $n=8$ the temperature exponents saturate to a constant value 
$\zeta^T_{\infty}\simeq 0.8$ \cite{CMV01}. 
The errorbars are estimated by the rms
fluctuations of the logarithmic slope. 
To ensure the statistical convergence of high-order moments
we collected three hundreds of snapshots of the fields, spaced by half 
of the large-eddy turnover time  $L/\langle {\bm v}^2 \rangle^{1/2}$.}
\label{fig:2} 
\end{figure}
Dimensional arguments based on this phenomenological picture would
lead to the Bolgiano-Obukhov scaling, in the range
 $l_B \ll r \ll L$:   
$S^T_n(r)=\langle [T({\bm r},t)-T({\bm 0},t)]^n \rangle \sim r^{n/5}$,
and
$S^v_n(r)=\langle 
[({\bm v}({\bm r},t)-{\bm v}({\bm 0},t))\cdot \hat{\bm r}]^n \rangle 
\sim r^{3n/5}$ (see, e.g., Ref.~\cite{S94} and references therein).
 Actually, due to the presence of structures of warm-rising or cold-descending
fluid -- the thermal plumes (see Fig.~\ref{fig:1}) -- the statistics of
temperature increments exhibits a nontrivial scale-dependence.
Indeed, as shown in Figure~\ref{fig:2}, moments of temperature
increments display a scaling behavior $ S^T_n(r) \sim r^{\zeta^T_n}$ 
characterized by exponents deviating from
the dimensional expectations. (Conversely, 
moments of velocity increments, 
$S^v_n(r)\sim r^{\zeta^v_n}$, do not
show measurable deviations from dimensional scaling, i.e. $\zeta^v_n=3n/5$.)
This {\em anomalous scaling} is
a feature shared by a large class of turbulent systems: understanding 
the origin of this phenomenon
from first principles is a major 
challenge of turbulence. With a significant exception. 
Indeed, for 
passive turbulent transport this problem has been recently solved. 
Let us briefly recall the main points, referring to 
Ref.~\onlinecite{FGV01} for a 
comprehensive review. We consider an idealized experiment of
turbulent dispersion of a passive tracer, e.g. dye, in a 
convective flow. 
The equation that governs the dynamics of the concentration of 
tracer is 
\begin{equation}
\partial_t C+{\bm v}\cdot {\bm \nabla} C =\kappa \Delta C + f_C \;,
\label{eq:2}
\end{equation}
while the velocity field ${\bm v}$ evolves according to Eq.~(\ref{eq:1}).
Although seemingly similar, the dynamics of temperature and
concentration fields are radically different:
temperature is an {\em active scalar}, since it affects
velocity via the buoyancy forces, whereas concentration is a
{\em passive scalar}. The concentration 
fluctuations show
an anomalous  scaling behavior 
$S^C_n(r)=\langle [C({\bm r},t)-C({\bm 0},t)]^n 
\rangle \sim r^{\zeta^C_n}$ as well: the
exponents $\zeta^C_n$ differ from the dimensional expectation.
Since $S^C_n(r)$ is a linear combination of
various $n$-point correlation function of the concentration field  
$\langle C({\bm x}_1,t) \ldots  C({\bm x}_n,t) \rangle$, the latter 
has to contain a contribution, denoted as 
$Z^C_n({\bm x}_1,\ldots,{\bm x}_n)$, that carries
the anomalous scale dependence. In mathematical terms, 
$Z^C_n(\lambda{\bm x}_1,\ldots,\lambda{\bm x}_n)=\lambda^{\zeta^C_n}
Z^C_n({\bm x}_1,\ldots,{\bm x}_n)$.
The main point is that the function $Z^C_n$ 
is characterized by a special dynamical property 
that distinguishes it from a generic 
scaling function. Let us remind 
the passive scalar equation~(\ref{eq:2}) can be written in the equivalent form
$\frac{d}{dt} C= f_C$, where $\frac{d}{dt}$ stands for the total derivative
along the particle trajectories defined by the 
stochastic differential equations
$d{\bm X} = {\bm v}({\bm X},t)dt+ \sqrt{2\kappa}\, d{\bm W}(t)$,
where ${\bm W}(t)$ is brownian motion.
The remarkable result is that
 $\frac{d}{dt} \langle Z^C_n \rangle_X= 0$,
where the total derivative is performed 
following $n$ particles advected by the flow, 
and the average is taken over the ensemble of all trajectories.
In plain words, $Z^C_n$ is {\em statistically preserved} by the flow
\cite{BGK98,CV01,ABCVP01}. In the specific context of a Gaussian, 
$\delta-$correlated velocity field -- the Kraichnan model --
this is equivalent to say that the functions $Z^C_n$ are zero-modes 
of the Fokker-Planck operator for $n$-particle diffusion   
\cite{GK95,CFKL95,SS95}. 
An important consequence of
statistically preserved structures
is that the passive scalar scaling exponents are
{\em universal} with respect to the choice of the injection $f_C$, 
since the latter does not enter the definition of $Z^C_n$.\\ 
We now turn our attention back to the temperature field.
What we have learnt in the passive scalar case suggests
to investigate the effect of the external forcing $f_T$ 
on the scaling exponents. 
In Fig.~\ref{fig:3} we show that 
the scaling exponents of temperature fluctuations are the same
for two different choices of injection terms $f_T$.
Therefore, we conclude that the exponents $\zeta^T_n$
are {\em universal} properties of two-dimensional Boussinesq convection.
The universality of scaling exponents suggests the possibility that a 
mechanism similar to that at work in passive scalar turbulence might be 
present in turbulent convection as well.
To further pursue this line of thought, we notice that in the case
of passive scalars, it is the whole function $Z^C_n$ to be universal
with respect to forcing, not only its scaling exponent. 
It is thus of interest to look at the anomalous part of the 
temperature correlation function 
$\langle T({\bm x}_1,t)\ldots T({\bm x}_n,t)\rangle$, 
to check whether it is universal. This measurement is unfortunately
quite difficult for two reasons.
First,
the correlation function depends on $2n$ independent coordinates;
even if we exploit the statistical symmetries of this function --
translational and scaling invariance -- and we limit ourselves to the 
isotropic contribution -- by averaging over all configurations differing only
by a rigid rotation -- there will still be $2n-4$ degrees of freedom; 
in the most favorable case, $n=4$
(for $n=2$ 
\begin{widetext}
\begin{figure*}[ht]
\includegraphics[draft=false, scale=1,clip=true]{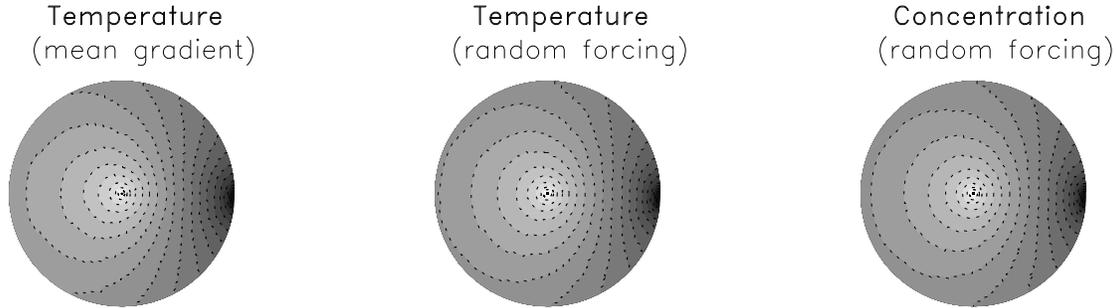}
\caption{The functions  $\sigma_T(r/R,\theta)=
S^T_{2,2}({\bm R},{\bm r})/S^T_4(R)$ (left and center),
and $\sigma_C(r/R,\theta)=
S^C_{2,2}({\bm R},{\bm r})/S^C_4(R)$ (right),
in polar coordinates $0<r/R<1$ and
$0<\theta<2\pi$, where $\theta=\cos^{-1}(\hat{\bm R}\cdot\hat{\bm r})$.
The color is white where the function is zero, black where it is equal
to unity. The function has a minimum at the origin, 
$\sigma(0,\theta)=0$, and a maximum  $\sigma(1,0)=1$ at ${\bm r}={\bm R}$.}
\label{fig:4} 
\end{figure*} 
\end{widetext}  
deviations from dimensional scaling are not detectable),
the configuration space has four dimensions, which makes it quite untractable.
Second,
the anomalous part of the correlation function is 
hidden among several other contributions: it can be extracted only by 
taking proper linear combinations, as, for example, in the case of
$S_n^T(r)$. To circumvent, at least partially, those problems,
we focus on a particular observable,
$S^T_{2,2}({\bm R},{\bm r})=
\langle [T({\bm R},t)-T({\bm 0},t)]^2 [T({\bm r},t)-T({\bm 0},t)]^2 \rangle$,
which is still anomalous, yet it has
a nontrivial geometrical content. 
Since for ${\bm r}={\bm R}$ it reduces to the usual $S_4(R)$ we 
can write its functional dependence as  $S^T_{2,2}({\bm R},{\bm r})=
S^T_4(R)\sigma_T(r/R,\theta)$ where  $\theta$ is the angle   
between the directions of ${\bm r}$ and ${\bm R}$.
Since the scaling exponent of $S^T_4(R)$ is universal, the bottom line is
whether the ``angular'' part $\sigma_T(r/R,\theta)$ is universal as well.  
In Fig.~\ref{fig:4} we show a plot of the function $\sigma$ for the
two different injection mechanisms (left and center).
The similarity between the two pictures points to the conclusion that
the anomalous part of the correlation function is again universal.\\ 
This result leads us to conjecture that 
statistically preserved structures
$Z^T_n({\bm x}_1,\ldots,{\bm x}_n)$ might exist also for temperature:
it is natural to define them by the property 
$\frac{d}{dt} \langle Z^T_n \rangle_X = 0$, as in the passive scalar case.
Notice however that, since temperature is an active scalar, this 
definition does not automatically ensure the universality of $Z^T_n$.
Indeed, even if the forcing does not appear explicitly in the definition
of $Z^T_n$, the statistics
of the trajectories ${\bm X}(t)$ in principle depends on $f_T$, via
the action of $T$ on ${\bm v}$. Therefore, should we accept
the existence of statistically preserved structures,
the universality of anomalous temperature correlations
requires to postulate that the whole 
statistics of ${\bm v}$ be universal as well.
In Fig.~\ref{fig:5} we show that this is indeed the case.\\
Statistically preserved structures for temperature
fluctuations entail another interesting consequence: 
since $Z^T_n$ is defined entirely in terms of the (universal) 
statistics of particle trajectories, and those are the same both
for temperature and for concentration, we expect that $Z^T_n=Z^C_n$.
In Fig.~\ref{fig:6} we show that the scaling exponents of 
temperature and concentration are equal, as expected. As 
for the ``angular part'' of $Z^T_n$, it is quite similar to
that of $Z^C_n$ (see Fig.~\ref{fig:4}, center and right).
This is a further indirect evidence for the existence of statistically
preserved structures for temperature statistics.\\
In conclusion, the global picture of scaling and universality 
in two-dimensional turbulent convection is as follows.
Velocity statistics is strongly universal with respect to
the external driving: probability density functions of
velocity fluctuations are self-similar, and close to a 
Gaussian distribution, independently of the choice of $f_T$.
This is most likely a consequence of the observed universal Gaussian behavior
of the inverse energy cascade in two-dimensional Navier-Stokes turbulence
\cite{PT98,BCV00}.
Indeed, velocity fluctuations in two-dimensional convection 
also arise from an inverse energy cascade which is
driven now by buoyancy forces. At variance with the usual Navier-Stokes
inverse cascade, the energy injection now is not 
restricted to small scales. Indeed, the energy input rate 
$\varepsilon(r)=
\beta {\bm g}\cdot\langle{\bm v}({\bm r},t)T({\bm 0},t)\rangle$
grows with the scale as 
$\varepsilon(r) \sim r^{4/5}$. This scale-dependent input
induces the observed scaling 
$S^v_n(r)\sim (\varepsilon(r) r)^{n/3} \sim r^{3n/5}$.
Temperature statistics shows anomalous scaling. This stems
from the existence of statistically preserved structures,
whose existence explains the observed universality of 
anomalous temperature correlation functions and the equality
between temperature and concentration anomalies.\\
Let us point out that the equivalence of the statistics 
of an active scalar, as temperature, to that of a passive scalar, as
concentration, depends crucially on the universality of the whole velocity
statistics found here. That could however be a nongeneric phenomenon 
in active scalar turbulence, and
depend on the specific form of the feedback of the scalar field on
the velocity. For example, in three-dimensional turbulent convection 
the Navier-Stokes equations are 
characterized by a direct and intermittent energy cascade,
the whole velocity statistics might then be nonuniversal, and
a new type of universality might emerge. \\
We acknowledge several useful discussions with R.~Benzi, L.~Biferale, S.~Toh. 
This work was supported by EU under the
contracts HPRN-CT-2000-00162 and FMRX-CT-98-0175.
A.~M. has been partially supported by the INFM project GEPAIGG01.
T.~M. acknowledges the Japan Scholarship Foundation.
Numerical simulations have been performed at IDRIS 
(projects 011226 and 011411), at CINECA (INFM Parallel Computing
Initiative), and at SX5 of the Yukawa Institute.

\end{document}